\documentclass[conference]{IEEEtran}
\IEEEoverridecommandlockouts
\usepackage{cite}
\usepackage{amsmath,amssymb,amsfonts}
\usepackage{algorithmic}
\usepackage{booktabs}
\usepackage{graphicx}
\usepackage{textcomp}
\usepackage{xcolor}
\usepackage{mathrsfs}
\usepackage{multirow}
\usepackage{color}
\usepackage{colortbl}
\usepackage{hyperref}
\def\BibTeX{{\rm B\kern-.05em{\sc i\kern-.025em b}\kern-.08em
    T\kern-.1667em\lower.7ex\hbox{E}\kern-.125emX}}
\begin{document}

\title{One Person, One Model—Learning Compound Router for Sequential Recommendation \\
}

\author{\IEEEauthorblockN{Zhiding Liu\textsuperscript{1,2},
Mingyue Cheng\textsuperscript{1,2}, Zhi Li\textsuperscript{3},
Qi Liu\textsuperscript{1,2}, Enhong Chen\textsuperscript{1,2,*}}
\IEEEauthorblockA{{\textsuperscript{1}Anhui Province Key Laboratory of Big Data Analysis and Application,} \\
{University of Science and Technology of China, Hefei, China} \\
{\textsuperscript{2}State Key Laboratory of Cognitive Intelligence, Hefei, China} \\
{\textsuperscript{3}Shenzhen International Graduate School, Tsinghua University, Shenzhen China} \\ 
{\textit{\{doge, mycheng\}@mail.ustc.edu.cn, \{qiliuql, cheneh\}@ustc.edu.cn , zhilizl@sz.tsinghua.edu.cn}}}
}
\maketitle
\begin{abstract}
Deep learning has brought significant breakthroughs in sequential recommendation (SR) for capturing dynamic user interests. A series of recent research revealed that models with more parameters usually achieve optimal performance for SR tasks, inevitably resulting in great challenges for deploying them in real systems. Following the simple assumption that light networks might already suffice for certain users, in this work, we propose CANet, a conceptually simple yet very scalable framework for assigning adaptive network architecture in an input-dependent manner to reduce unnecessary computation. The core idea of CANet is to route the input user behaviors with a light-weighted router module. Specifically, we first construct the routing space with various submodels parameterized in terms of multiple model dimensions such as the number of layers, hidden size and embedding size. To avoid extra storage overhead of the routing space, we employ a weight-slicing schema to maintain all the submodels in exactly one network. Furthermore, we leverage several solutions to solve the discrete optimization issues caused by the router module. Thanks to them, CANet could adaptively adjust its network architecture for each input in an end-to-end manner, in which the user preference can be effectively captured. To evaluate our work, we conduct extensive experiments on benchmark datasets. Experimental results show that CANet reduces computation by $55\sim65\%$ while preserving the accuracy of the original model. Our codes are available at https://github.com/icantnamemyself/CANet.
\end{abstract}

\renewcommand{\thefootnote}{}
\footnotetext{\textsuperscript{*} Enhong Chen is the corresponding author.}
\renewcommand{\thefootnote}{\arabic{footnote}}

\section{Introduction}
Recent years have witnessed the rapid growth of the Internet, resulting in the explosive information. To overcome the information overload issue, recommend systems have been proposed to mine user preferences to help them find their interested items \cite{liu2011personalized, huang2022personal}. Among them, sequential recommendation (SR) models have become ever increasingly popular in capturing evolving interests by modeling temporal dependence of user historical behaviors. A large body of algorithms have been proposed and perform well in SR tasks \cite{fpmc, highorder}.

Recently, deep learning-based SR models have achieved great breakthroughs since these methods~\cite{li2018learning, sasrec, srgnn} usually exhibit more expressive capacity in modeling sequential user interactions than previous ones. Generally, these deep SR models are composed of several middle layers and two embedding layers. A widely accepted view is that recommendation models are generally huge, and several recent efforts have been devoted to reveal that a larger network could yield obvious performance improvements in SR tasks \cite{nasr, stackrec}. Nevertheless, we argue that such improvements are unaffordable in real recommender systems since larger models inevitably result in expensive computational cost and require higher-performance hardware. A well-known scene is to allow many mobile smartphones to be equipped with SR models so as to provide personalized online services, in which very limited computation costs can be leveraged for recommendation. Hence, one question naturally arises that how can we improve the efficiency of these large SR models by saving the utilization of hardware resources while maintaining the overall performance.

\begin{figure}
	\centering
	\includegraphics[width=\linewidth]{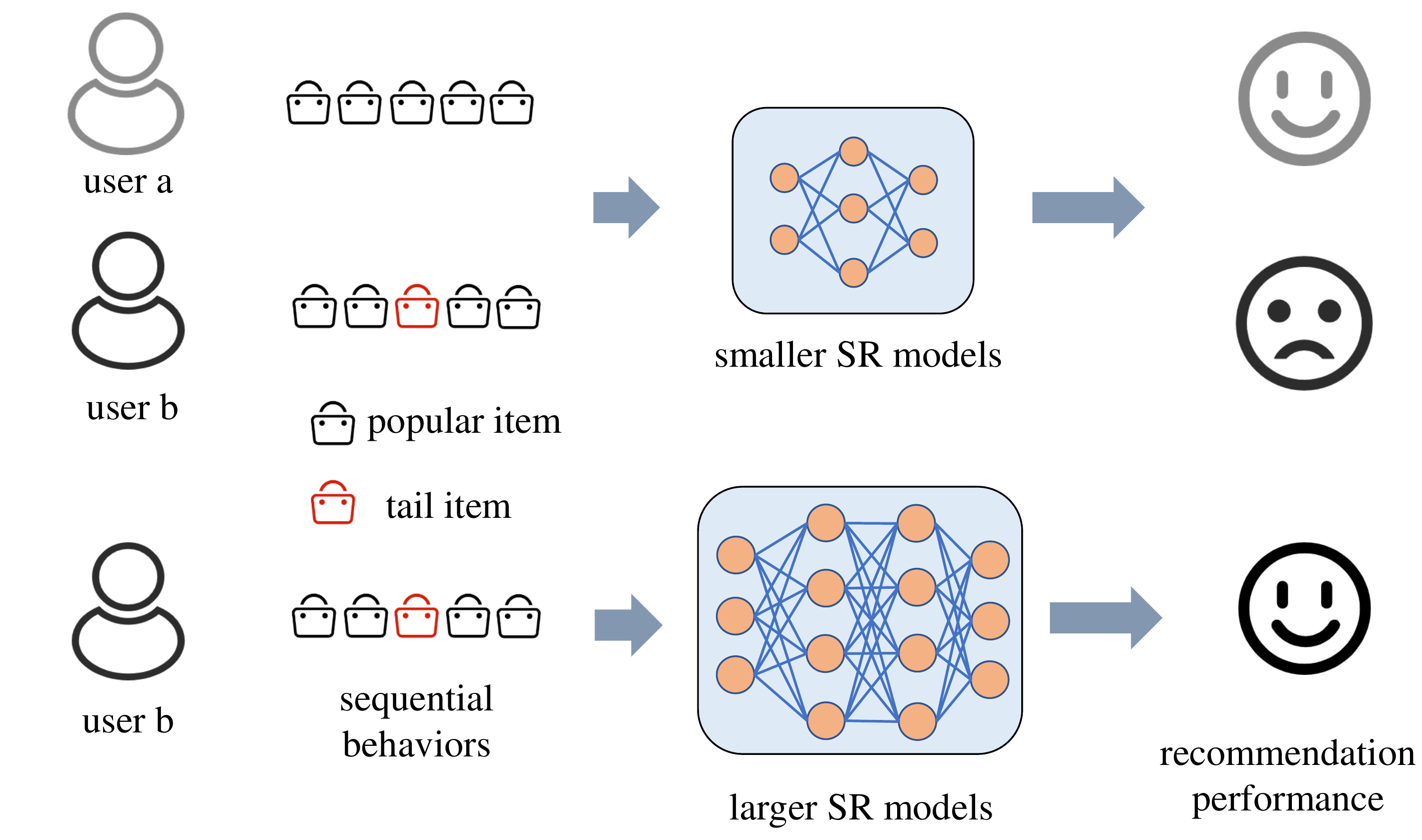}
	\vspace{-0.3in}
	\caption{Recommendation results for different users in a long-tail view.}
	\vspace{-0.2in}
	\label{fig:user}
\end{figure}

Actually, some attempts have been provided in recent researches. Following the traditional one-model-fits-all paradigm, some pioneer works propose to compress heavy models to light models by knowledge distillation and matrix decomposition \cite{icdm_kd, fan2021lighter, compression}. Besides, embedding size search methods \cite{nis,autodim} try to figure out a suitable embedding size for input features to achieve storage space efficiency in a automated schema \cite{zheng2022automl}. Further in a view of dynamic architecture, SkipRec \cite{skiprec} utilizes the layer-skipping schema to reduce unimportant computation in the middle layers. Despite their effectiveness, we hold these methods are not scalable enough since they implement a fixed network or a single-dimensional adaptive architecture. Different from these works, we aim to explore a new research question, i.e., equipping each person with one adaptive backbone network in a compound view. To be more specific, our goal is to design a model-agnostic framework which adaptively assigns network architectures for each user in multiple aspects of network hyper-parameters. We refer to such a goal as One person, One model.

Our solution is oriented in a in-depth thought about the personalities of users: larger and deeper SR models are not always necessary for all users. As illustrated in Figure \ref{fig:user}, although larger networks can show more promising results, such settings might be rather heavy and burdensome for certain users. To evidence our thought, we provide interesting quantitative empirical findings by conducting extensive experiments on benchmarks. As shown in section \ref{Empirical Studies}, we find that: (1) Larger models usually require several times of the computational cost than smaller ones to achieve improvements, it is essential to ensure the efficiency of large models; (2) Distinguishing all users according to the existence of long-tail item distributions, results show that smaller models can perform well enough for users who only interested in popular items.

Guided by the above analysis, in this paper, we propose a simple yet very scalable framework for SR tasks, namely \textbf{C}ompound \textbf{A}daptive \textbf{Net}work or \textbf{CANet} in short. Specifically, we first construct the routing space with various submodels parameterized by their sizes like embedding size and layer depth. Such method is competent to tackle difficult heterogeneous searching of compound parameter space and is also flexible enough to suit homogeneous tasks like model depth searching. While to avoid extra storage overhead of the routing space, we utilize weight-slicing schema \cite{slim,dsnet} to maintain all the submodels in exactly one supernet. Besides, we introduce Gumbel-Softmax \cite{gumbel} in our framework to derive the non-derivable problem caused by discrete arg-max operation in routing, making the training process end-to-end. Moreover, we design simple yet powerful auxiliary optimization objectives, which help to train all the submodels fairly and guarantee the router evolves correctly. Finally, we evaluate the effectiveness and inference efficiency in three benchmark datasets, results show that our proposed framework can reduce up to 65\% computation cost with minimal degradation (less than 1.5\%) in recommendation performance.

Our contributions are summarized as follows:
\begin{itemize}
\item We argue that the one-model-fits-all paradigm brings computation redundancy in recommendation. For example, a light model may suffice for users who only interested in popular items. It is of practical significance to study how to build a personalized recommendation model to achieve a win-win situation of effectiveness and efficiency.
\item We propose CANet, a novel framework which adaptively adjusts its architecture with respect to input user sequence to reduce redundant computaion while maintaining the good performance of large models. 
\item We conduct sufficient experiments with prevalent SR recommenders on three datasets. The results demonstrate that our method can significantly reduce the computation with a slight degradation of recommendation results. 
\end{itemize}

\section{Preliminaries}
We first formalize the sequential recommendation (SR) task and introduce the notations. Then we briefly introduce the deep SR architecture and present a linear residual modification. After that, we conduct insightful empirical studies, which is the core motivation of our work.

\subsection{Definitions and Notations}

Given the user set $U=\{u_1,u_2,...,u_{|U|}\}$ with $|U|$ unique users and item set $I=\{i_1,i_2,...,i_{|I|}\}$ with $|I|$ unique items, the most basic task of SR models is to predict the next item $x_{t+1}^u \in I$ which user $u$ will interact with based on user's past behaviours $X^u=[x_1^u,x_2^u,...,x_t^u]$, where $x_t^u$ indicates the $t-th$ interacted item of user $u$. The key idea under SR models is to mining the dependency of user-item sequences so as to predict the next item.

\subsection{A Review on Deep Sequential Recommendation Models}
\subsubsection{Basic Models}
The most prevalent deep SR models are always composed of several middle layers and two embedding layers. To be specific, each input item $x_i^u$ is first mapped into embedding space through the item embedding layer, then the discrete user-item interactions can be represented by an embedding matrix $E^u=[e_1^u, e_2^u,...,e_t^u]$. Next, the SR model feeds the embedding matrix into its middle layers to extract the dependence of sequential behaviors. Finally, the extracted hidden representations are fed to the next item prediction layer to generate user interests over candidate items. 

\subsubsection{Learnable Residual Connection}
By revisiting recent deep SR models \cite{deep_sr_survey}, we notice that most current deep SR models can only stack shallow layers \cite{sasrec,bert4rec,clue} by residual learning due to the difficulty of training deeper networks \cite{residual}. The shallow architectures limit their capabilities significantly. To solve the problem, taking inspiration from recently proposed works \cite{rezero, nasr}, we add a learnable parameter $\lambda_L$ initialed as 0 to each skip connection component. Specifically, for the SASRec \cite{sasrec} backbone contained by CANet, the transformer architecture enhanced with our simple residual modification can be expressed as
\begin{equation}
H_L = LN(H_{L-1}+\lambda_L*sublayer_{L-1}(H_{L-1}))
\end{equation}
where $LN(\cdot)$ indicates the layer-norm \cite{ln} operation and sublayer is in between \textit{Multi-head Self-attention} layer and \textit{Point-wise Feed-forward} layer. 

\subsection{Empirical Studies}\label{Empirical Studies}
Considering various models' popularity and performance, we implement SASRec with different configurations as our base models and evaluate their performance by Normalized Discounted Cumulative Gain (NDCG@10) on Last.FM20 dataset with controlled parameters. We also evaluate the computation cost by Floating-Point Operations\footnote[1]{https://github.com/Lyken17/pytorch-OpCounter} (FLOPs) of the models to inference on an input user sequence. Furthermore, we split items into short-head items and long-tail items by their popularities. Similar to \cite{icdm_tail}, we regard the top 20\% as head items and the rest are treated as tail items. By regarding users who have interacted with the long-tail items as `tail users' and the rest as `head users', we can further study the recommendation results for users with diverse preferences.

\begin{figure}
	\centering
	\includegraphics[width=\linewidth]{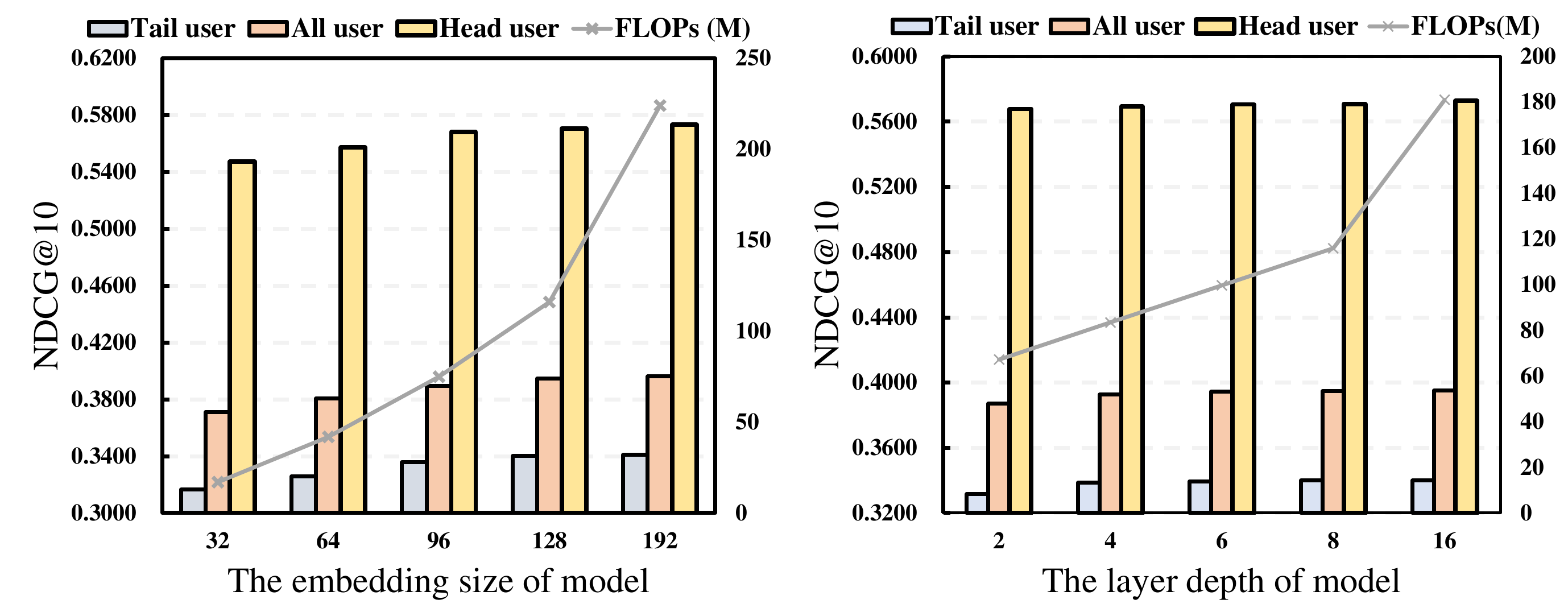}
	\vspace{-0.3in}
	\caption{Recommendation performance to different users and computation cost w.r.t. the width and depth of SASRec tested on Last.FM20}
	\vspace{-0.2in}
	\label{fig:motivation}
\end{figure}

To keep the comparison fair, the experimental setting strictly obeys the hyper-parameter rules. When comparing the effect of embedding size, all the models are trained with 8 layers. Similarly, models are all trained with an embedding size of 128 when comparing the effect of layer depth. With rigorous controlled experiments, we report all the results by (1) adopting auto-regressive regression accompanied by cross-entropy loss as an optimization objective, (2) generating the probability of all items without leveraging shared item embedding layer during training and evaluation, and (3) training all SR models until achieving the best results. We report all experimental results in Figure \ref{fig:motivation}.

By analyzing the experimental results, we can draw conclusion as follows:
\begin{itemize}
\item Through learnable residual connections, deep learning based SR models can scale up to larger models to a certain extend, and the recommendation performance of the model grows consistently with the size. Notably, the largest model obtains 6.7\% improvement over the narrowest one and 2.1\% improvement over the shallowest one on the Last.FM20 dataset. 
\item The FLOPs grow nearly linearly with the expansion of embedding size and the increase of middle layers. Consequently, models should pay much more computation (13.21x and 2.69x respectively) to achieve the abovementioned improvements.
\item Small models can perform well enough for users who only interested in popular item. Besides, the performance for tail users gains more (8.1\% and 2.7\% respectively) than average from large models. 
\end{itemize}

To summarize, the empirical studies show that larger SR models can perform better but require multiple computational cost. Besides, we also observe that some user preferences can be captured by small models while others require more computation. However, existing models usually execute in a static way, which may introduce redundant computation. These observations naturally motivate us to think whether can we model user's interests by personalized architectures to reduce the redundancy in large models. Combined with the widely studied dynamic network technique \cite{dynamic_survey}, we would introduce our CANet in the next section.

\section{Framework of CANet}
In this section, we describe CANet in detail, which assigns suitable submodels to input user sequences adaptively to reduce computation cost with minimal degradation on recommendation performance. To the best of our knowledge, most previous works are focused on designing a general powerful architecture manually or automatically for all the inputs, while only a few attempts are devoted to modeling user preferences by personalized architectures \cite{skiprec}. We first present the whole framework then introduce the optimization strategies.

\subsection{Framework Description}\label{FD}
\begin{figure*}[htbp]
	\centering
	\includegraphics[width=\linewidth]{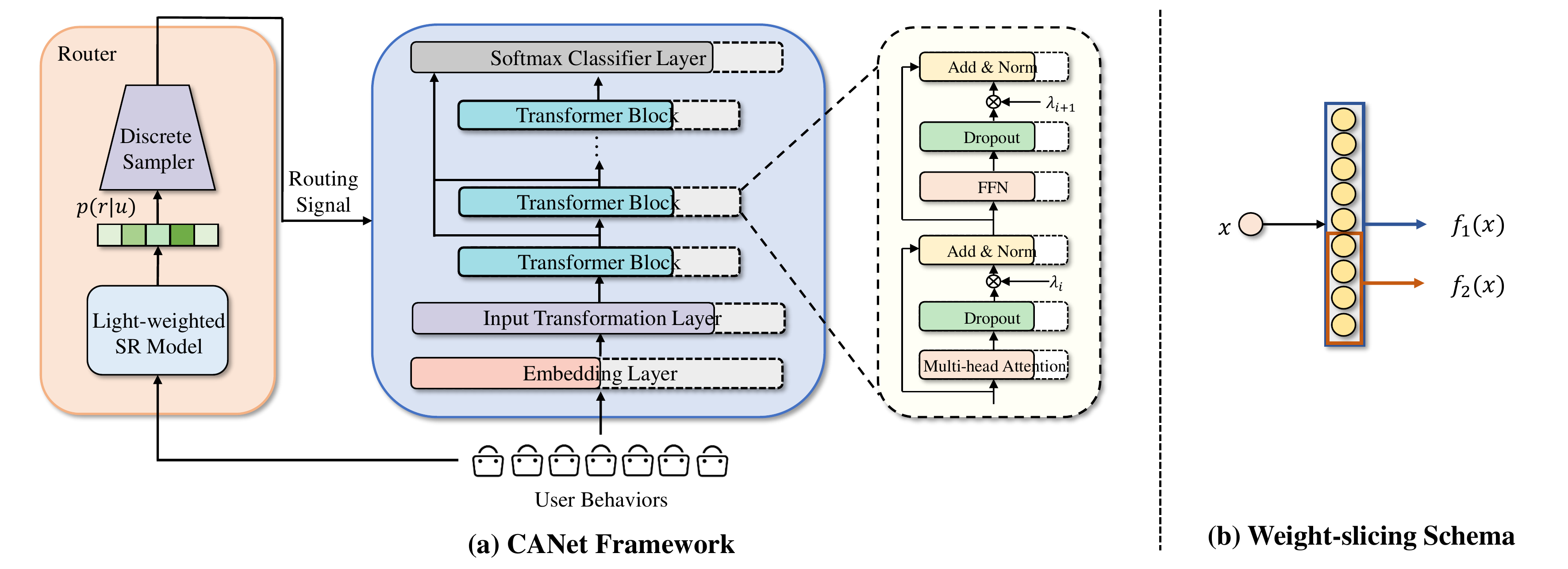}
	\vspace{-0.3in}
	\caption{Illustration of CANet framework and weight-slicing schema. Embedding size, hidden size and layer depth of the backbone network are adjusted adaptively according to the input user behaviors.}
	\vspace{-0.2in}
	\label{fig:framework}
\end{figure*}

As illustrated in Figure \ref{fig:framework}(a), our proposed CANet framework is made up of a light-weighted router and a backbone network. Since we empirically find that both enlarging embedding size and stacking more layers contribute to a better recommendation performance in section \ref{Empirical Studies}, existing works that try to figure out proper layers or assign fit embedding size for the input may lead to a sub-optimal result as all of them only implement a pure single-dimensional adaptive architecture \cite{dynabert}. A natural solution is to build a compound network space where more structural parameters can be considered. To be more specific, we need to design a model-agnostic framework to adaptively assign network architectures for each user in multiple aspects of network hyper-parameters.

\subsubsection{Router}\label{router section}
In practice, we regard each submodel as a unique path. In the meantime we train the light-weighted router learning to predict the suitability of routing user behaviors to corresponding submodels $p(r|u)$ in a user-specific manner. The chosen of path for user $u$ can be described as following discrete operation:
\begin{equation}
R_u = \arg\max\limits_{r} p(r|u)\label{argmax}
\end{equation}

The router is made up of a sequential feature extractor and a discrete sampler, responsible for generating discrete routing signals based on input user behaviors. The sequential feature extractor takes user interaction histories as input and outputs the probabilities $p(r|u)$ over all routes with respect to user $u$ adaptively. Since its goal can be seen as a task of predicting the user's "preference" for submodels in the routing space, any SR model can serve as a sequential feature extractor without loss of generality. In our experiment, to ensure the capability of our router and avoid unnecessary computation, we implement it using a single layer SASRec \cite{sasrec} model with an embedding size of 32. As for the discrete sampler, it performs sampling according to the probability $p(r|u)$ to generate the final routing signals. We employ Gumbel-Softmax \cite{gumbel} in our sampler to guarantee the differentiability of discrete operations, which will be discussed in detail in section \ref{OS}.

\subsubsection{Backbone Network}
The router tries to figure out a suitable submodel while the backbone network is focused on modeling user preference, which contains the whole routing space and adjusts its architecture according to the routing signals. We also instantiate a SASRec model as our backbone network. Notably, the whole framework is not limited to self-attentive-based backbones, it is potentially applicable for more SR models when the corresponding dynamic computation as Eq. \eqref{dynamic linear} and Eq. \eqref{dynamic ln} is well defined. As for more exploration of competitive backbones like TCN-based networks \cite{nextitnet}, we leave them for future work.

To cover a compound parameter space, we implement the routing space made up of three aspects: embedding size, hidden size and depth. Embedding size implies the original width of item embedding while hidden size stands for the dimension of the middle layers. Then the whole routing space can be described by the product of these parameters. It is worth noting that our design can well generalize to existing works \cite{skiprec,autodim} by only considering layer depth or embedding size in the routing space. Moreover, it can scale up to more complex parameters like layer-wise hidden size instead of being limited to the experimental setup.

The framework can explore a huge space by such definition. Assume that there are $a$ candidate choices for embedding size and $b,c$ choices for hidden size and depth respectively, then the routing space of each user consists of $n=a\times b\times c$ candidate architectures. The numerous submodels can be a large storage overhead, which is inappropriate for reducing redundancy in large models. To tackle this challenge, combining the fact that these submodels share a similar structure, we utilize the weight-slicing schema \cite{slim} and thus arrange all the candidate submodels in one supernet.

Specifically, in the weight-slicing schema, the computation of a general layer can be well described as in Figure \ref{fig:framework}(b), where $f_1(x)$ indicates a full transformation while $f_2(x)$ stands for a computation with partial parameters. Taking inspirations form recent works \cite{dsnet}, here we treat parameters as sequences and force all the sub-parameters of submodels to begin from the first index of full parameters. Formally, for the linear transformation operation with a transformation matrix $W$ and bias $b$, supposing its input and output dimension are $D_{in}, D_{out}$ respectively, the full transformation and transformation with input and output dimension of $k_1<D_{in}, k_2<D_{out}$ can be computed by the following formula:

\begin{equation}\label{dynamic linear}
\left\{
\begin{aligned}
f_1(x) & = Wx + b &\\
f_2(x) & = W[:k_2, :k_1]x + b[:k_2] &\\
\end{aligned}
\right.
\end{equation}

Besides, with respect to the layer-norm operation used in SASRec backbone with weight $W$ and bias $b$ at an input dimension of $D_{in}$, the dynamic computing can also described as follows:
\begin{equation}\label{dynamic ln}
\left\{
\begin{aligned}
f_1(x) & = LN(W,b,x) &\\
f_2(x) & = LN(W[:k],b[:k],x) &\\
\end{aligned}
\right.
\end{equation}
where $LN(\cdot)$ indicates the layer-norm function and $k<D_{in}$ corresponds to the input dimension of $f_2(x)$. 

Equipped with the weight-slicing schema, the backbone network contains exactly one largest model serving as the supernet. Any smaller model can be indexed in the supernet by weight slicing and early stopping methods, thus the router can search in a wide searching space while introducing no additional storage overhead. On the basis of SASRec, we modify the linear transformation operation used in \textit{Multi-head Attention} layer and \textit{FFN} layer into its dynamic version defined by Eq. \eqref{dynamic linear}, and the layer-norm operation correspondingly. We also insert an extra dynamic linear layer serving as the \textit{Input Transformation Layer}, which convert the item embeddings to the dimension of the middle blocks. Besides, we add shortcuts to \textit{Softmax Classifier} for the outputs of each block, achieving a reduction in depth by early stopping. Note that we do not add internal classifiers like \cite{over-think,deebert} do but share one classifier among all shortcuts. This is because the classifier always tends to be very huge due to the large item set, which will bring much parameter overhead.

\subsection{Optimization Strategies}\label{OS}
\subsubsection{Gumbel-Softmax}
The choice of routes is in a hard selection schema as shown in Eq. \eqref{argmax}, making the framework not end-to-end differentiable. Meanwhile the arg-max operation will ignore the probabilistic significance, which means the results of multiple sampling always be the same but not be consistent with the probability distribution. To tackle these challenges, we introduce the Gumbel-Softmax trick \cite{gumbel} in our discrete sampler, which simulates the non-differentiable discrete sampling from a categorical distribution to approximate the hard selection over routes. Though such method, we can ensure the sampling results consistent to the predicted distribution while maintaining the gradients.

Formally, suppose weights $p(r|u)=\{p_1^u,p_2^u,...,p_n^u\}$ are the normalized probabilities of $n$ routes with respect to user $u$, then the chosen of path for user $u$ enhanced by Gumbel-Max trick can be drawn as:

\begin{equation}
R_u = \arg\max\limits_{i\in [1,n]} (log p_i^u + g_i)\label{gumble-max}
\end{equation}

The gumbel noises $\{g_1,g_2,...,g_n\}$ are i.i.d. samples drawn from $Gumbel(0,1)$ distribution, computed by $g_i = -\log(-\log(a))$ where $a$ is sampled from a uniform distribution, i.e., $a \sim Uniform(0,1)$. With gumbel noises perturbing the logits $\{\log p_i^u\}$, the arg-max operation is equivalent to drawing a sample from probabilities of $p(r|u)$. By this approach, We first solve the probability agnostic problem of discrete sampling. To further alleviate the non-differentiable issue, we deploy the Gumbel-Softmax trick as a continuous relaxation to arg-max using:
\begin{equation}
\alpha _i^u = \frac{\exp{(\frac{\log p_i^u+g_i}{\tau})}}{\sum _{j=1}^n\exp{(\frac{\log p_j^u+g_j}{\tau})}} \label{gumbel-softmax} 
\end{equation}
where $\tau$ is the temperature parameter to control the discreteness of the output, we simply set it at 1 by default.

During the training process, we sample discrete routes according to Eq. \eqref{gumble-max} and estimate the gradients of the sampling operation through Eq. \eqref{gumbel-softmax}. By using the above approach, the router's parameters can receive well-defined gradients, consequently the router and the backbone network can be optimized simultaneously in an end-to-end manner without introducing more bias in 2-stage training.

\subsubsection{Fairness in Supernet}
As discussed above, the routing space of CANet is a supernet consisting of all the submodels, alleviating the challenge of storage overhead while leaving the model collapse issue \cite{dsnet} aside. That is, supposing an unbalanced initial predicted distribution of the router, which leads one of the submodels to get more supervised training signals from the recommendation task. Therefore this submodel will converge much faster than others, resulting in more attention paid from the router. Ultimately, the router will collapse into a static one without learning any user features to decide on routing signals.

To tackle the trouble, we add an auxiliary loss to encourage training fairness in the supernet. Inspired by \cite{switch}, we propose to construct an uniform signal $p_{\rm{uni}}=[\frac{1}{n},\frac{1}{n},...,\frac{1}{n}]$ as a soft target and apply Cross-Entropy loss on the predicted distribution, forming the Uniform loss as follows:

\begin{equation}
L_{\rm{Uni}} = L_{\rm{CE}}(p(r|u), p_{\rm{uni}})
\end{equation}

Such auxiliary loss can force the predicted distribution to be closer to a uniform distribution, ensuring that each submodel in the supernet can receive enough supervised training signals. Experimental results show that with Uniform loss combined, we can well avoid the collapse of the router.

\subsubsection{Prior Guide}
When only using the Gumbel-Softmax and the Uniform loss above, the optimization process of the router is a black box thus the routing results may lack of interpretability or even be unreliable. Therefore, it is natural to encode the prior knowledge found in empirical studies into the training process, serving as a guidance signal to guarantee the router evolves correctly.

In detail, we first use the smallest submodel, i.e., the model with the smallest embedding size, hidden size and least layers, to predict the user's last interacted item for each training sequence. Then we evaluate the prediction results with $Recall@k$, where $k$ is a hyper-parameter. In this way, we jointly consider the efficiency and effectiveness of the model, and further we can verify users into two classes according to the evaluation: (1) \textbf{`Easy users'} $u_{\rm{easy}}$ that the smallest model can well predict their preferences, corresponding to $Recall@k$ equals to 1; (2) \textbf{`Hard users'} $u_{\rm{hard}}$ that more computation is required to make a good recommendation, corresponding to $Recall@k$ equals to 0. 

Intuitively, modeling hard users and easy users separately with large and small models can reduce computation cost while maintaining recommendation performance. We generate exponential distributions with peaks at the minimum parameter $y_{\rm{emb}}(u_{\rm{easy}}),y_{\rm{hidden}}(u_{\rm{easy}}),y_{\rm{layer}}(u_{\rm{easy}})$ as soft labels for easy users with respect to embedding size, hidden size and layer depth. Similarly, we generate exponential distributions with peaks at the maximum parameter $y_{\rm{emb}}(u_{\rm{hard}}),y_{\rm{hidden}}(u_{\rm{hard}}),y_{\rm{layer}}(u_{\rm{hard}})$ as soft labels for hard users. Based on the generated soft target, we define the Guide loss as the Cross-Entropy between the predicted probabilities and the generated guidance signals:
\begin{equation}
\begin{aligned}
L_{\rm{Guide}} &=L_{\rm{CE}}(p(emb|u), y_{\rm{emb}}(u))\\
&+L_{\rm{CE}}(p(hidden|u), y_{\rm{hidden}}(u))\\
&+L_{\rm{CE}}(p(layer|u), y_{\rm{layer}}(u))
\end{aligned}
\end{equation}
where $p(emb|u)$ stands for the predicted distribution on different embedding size, and so on for the others. $p(\cdot|u)$ can be computed by summing up the probabilities in $p(r|u)$ of routes with corresponding parameters.

The Guide loss works in two aspects: (1) For the easy users who can be well modeled by the smallest submodel, it pushes them to pass through smaller submodels to reduce computation redundancy. (2) For the hard users, the loss encourages them to pass through larger submodels in order to perform better results. In this way the prior knowledge can be novelly encoded in the training process, meanwhile the framework can produce reasonable routing signals based on the prior guidance.

Combined with all the optimization strategies above, the router and the backbone network can be trained jointly in an end-to-end manner. Supposing $L_{\rm{SR}}$ is the next-item prediction loss, then the optimization objective of our CANet framework can be formulated as following multi-task learning:

\begin{equation}
L =L_{\rm{SR}}+\lambda _1 L_{\rm{Uni}}+\lambda _2 L_{\rm{Guide}}\\
\end{equation}
where $\lambda_1$ and $\lambda_2$ are trade-off factors.

\section{EXPERIMENTS}
\subsection{Experimental Setup}
\subsubsection{Datasets Description}
To verify the effectiveness of our proposed CANet, we conduct extensive experiments on three real-world datasets: Last.FM20, Last.FM50 and RESS. In the data processing step, to alleviate the impact of cold items, we perform the basic pre-processing by filtering out interactions with less than 5 users. We summarize the statics of datasets in Table \ref{statistics}.
\begin{itemize}
\item Last.FM20 and Last.FM50\footnote[2]{http://www.dtic.upf.edu/ocelma/MusicRecommendationDataset/lastfm-1K.html}: We define the session length as 20 or 50 respectively, and extract successive items with corresponding length as the input sequence. We ignore sessions in which the time span between the last two items is longer than 2 hours.
\item RESS\footnote[3]{https://www.kaggle.com/datasets/mkechinov/ecommerce-behavior-data-from-multi-category-store}: This dataset contains seven months of user session from a multi-category online store \cite{transformers4rec}. For training efficiency, we randomly sampled 500,000 users and only use the purchase event from the month of Oct. 2019 with a max length of 20. Sequences shorter than 20 will be padded with zero at the beginning of the sequence.
\end{itemize}

\setlength{\tabcolsep}{3mm}{
\begin{table}[hbp]
  \centering
  \vspace{-0.1in}
  \caption{Statistics of the datasets}
    \begin{tabular}{c|cccc}
    \toprule
    DATA & \# actions & \# sequeces & \# items & length \\
    \midrule
    Last.FM20 & 10,699,640 & 534,892 & 199,013 & 20 \\
    Last.FM50 & 10,664,150 & 213,283 & 198,802 & 50 \\
    RESS & 4,418,277 & 500,000 & 112,705 & 20 \\
    \bottomrule
    \end{tabular}%
  \vspace{-0.1in}
  \label{statistics}%
\end{table}}%

\subsubsection{Routing Space and Framework Setup}
The CANet framework performs adaptive routing in a supernet based on SASRec. The supernet consists of 8 transformer blocks with multi-head number of 4 along with a max embedding size of 128. We split the routing space into three aspects, i.e., embedding size, hidden size and layer depth, where embedding size and hidden size represent the width of the embedding layer and the width of the middle layers. The detailed routing space is shown in Table \ref{routing space}, indicating the 36 possible submodels in the supernet. For experiments on three datasets, we set $k$ in prior guide as 5, 5, 10 respectively and $\lambda _1 =\lambda _2 = 0.01$.

\begin{table}[htbp]
  \centering
  \vspace{-0.15in}
  \caption{Routing Space}
    \begin{tabular}{l|c}
    \toprule
    Embedding Size  & {64, 96, 128} \\
    Hidden Size & {64, 96, 128} \\
    Layer Depth & {2, 4, 6, 8} \\
    \bottomrule
    \end{tabular}%
  \vspace{-0.15in}
  \label{routing space}%
\end{table}%

\subsubsection{Compared Methods}
To evaluate the effectiveness of CANet, we prepare several baselines to compare with our design. To keep the results of all recommendations as fair as possible, we strictly obey the hyper-parameter control rules and adopt auto-regressive regression accompanied by cross-entropy loss as optimization objective to train these deep SR models. In addition, residual-based SR models are enhanced with the learnable residual connection.
\begin{itemize}
\item \textbf{FPMC \cite{fpmc}} is a traditional SR model. The method utilizes matrix factorization to model user's long-term interest and predicts user's short-term preference base on Markov Chain. Following original settings, we use BPR loss to train the model.
\item \textbf{GRU4Rec \cite{gru4rec}} is a well-known session-based recommendation method by using the hidden state to memorize user's dynamic interests to the evolving of sequential behaviors. The number of gru layers is set to 2.
\item \textbf{SASRec \cite{sasrec}} is a self-attentive-based sequential recommendation model, which uses transformer structure to capture user's evolving preference based on their chronological behaviors. 
\item \textbf{NextItNet \cite{nextitnet}} employs the residual block of dilated convolution to increase the reception field for modeling user's long-term preference. We use 8 blocks with dilated factor of \{1, 4\} as our baseline model.
\item \textbf{SkipRec \cite{skiprec}} is a user-specific depth selection framework where the number of network layers can be selected on a per-user basis. The max searching depth is 8.
\item \textbf{AutoDim \cite{autodim}} is a framework that automatically select dimensions for input features. Here we treat each item as an unique feature and the searching space of embedding size is set to \{64, 96, 128\}.
\end{itemize}

To better illustrate the comparison of the recommendation performance, here we use SASRec-large and SASRec-small to denote model with the same size as CANet's backbone and model with similar computation cost to CANet respectively. Here we do not compare CANet with traditional compression methods \cite{icdm_kd, fan2021lighter,compression}, as our model can be built on top of them, the two do not conflict.

\setlength{\tabcolsep}{2mm}{
\begin{table*}[htbp]
  \centering
  \caption{Performance comparasion in terms of NDCG@10, Recall@10 and FLOPs}
    \begin{tabular}{p{5.94em}|c|ccc|ccc|ccc}
    \toprule
    \multirow{2}[2]{*}{\textcolor[rgb]{ .2,  .2,  .2}{\textbf{Model}}} & \multicolumn{1}{c|}{\multirow{2}[2]{*}{\textcolor[rgb]{ .2,  .2,  .2}{\textbf{Dynamic}}}} & \multicolumn{3}{c|}{\textbf{Last.FM20}} & \multicolumn{3}{c|}{\textbf{Last.FM50}} & \multicolumn{3}{c}{\textbf{RESS}} \\
    \multicolumn{1}{c|}{} &    & \multicolumn{1}{p{4.2em}}{\textcolor[rgb]{ .2,  .2,  .2}{\textbf{NDCG@10}}} & \multicolumn{1}{p{4.2em}}{\textcolor[rgb]{ .2,  .2,  .2}{\textbf{Recall@10}}} & \multicolumn{1}{p{5em}|}{\textcolor[rgb]{ .2,  .2,  .2}{\textbf{FLOPs (M)}}} & \multicolumn{1}{p{4.2em}}{\textcolor[rgb]{ .2,  .2,  .2}{\textbf{NDCG@10}}} & \multicolumn{1}{p{4.2em}}{\textcolor[rgb]{ .2,  .2,  .2}{\textbf{Recall@10}}} & \multicolumn{1}{p{5em}|}{\textcolor[rgb]{ .2,  .2,  .2}{\textbf{FLOPs (M)}}} & \multicolumn{1}{p{4.2em}}{\textcolor[rgb]{ .2,  .2,  .2}{\textbf{NDCG@10}}} & \multicolumn{1}{p{4.2em}}{\textcolor[rgb]{ .2,  .2,  .2}{\textbf{Recall@10}}} & \multicolumn{1}{p{5em}}{\textcolor[rgb]{ .2,  .2,  .2}{\textbf{FLOPs (M)}}} \\
    \midrule
    \textcolor[rgb]{ .2,  .2,  .2}{FPMC} & \textcolor[rgb]{ .2,  .2,  .2}{× } & \textcolor[rgb]{ .2,  .2,  .2}{0.3655 } & \textcolor[rgb]{ .2,  .2,  .2}{0.4345 } & \textcolor[rgb]{ .2,  .2,  .2}{101.894} & \textcolor[rgb]{ .2,  .2,  .2}{0.3675 } & \textcolor[rgb]{ .2,  .2,  .2}{0.4472 } & \textcolor[rgb]{ .2,  .2,  .2}{101.786 } & \textcolor[rgb]{ .2,  .2,  .2}{0.1284 } & \textcolor[rgb]{ .2,  .2,  .2}{0.2385 } & \textcolor[rgb]{ .2,  .2,  .2}{57.704} \\
    \textcolor[rgb]{ .2,  .2,  .2}{GRU4Rec} & \textcolor[rgb]{ .2,  .2,  .2}{× } & \textcolor[rgb]{ .2,  .2,  .2}{0.3887 } & \textcolor[rgb]{ .2,  .2,  .2}{0.4496 } & \textcolor[rgb]{ .2,  .2,  .2}{67.926} & \textcolor[rgb]{ .2,  .2,  .2}{0.4090 } & \textcolor[rgb]{ .2,  .2,  .2}{0.4697 } & \textcolor[rgb]{ .2,  .2,  .2}{123.342 } & \textcolor[rgb]{ .2,  .2,  .2}{0.1507 } & \textcolor[rgb]{ .2,  .2,  .2}{0.2760 } & \textcolor[rgb]{ .2,  .2,  .2}{57.832} \\
    \textcolor[rgb]{ .2,  .2,  .2}{Nextitnet} & \textcolor[rgb]{ .2,  .2,  .2}{× } & \textcolor[rgb]{ .2,  .2,  .2}{0.3863 } & \textcolor[rgb]{ .2,  .2,  .2}{0.4473 } & \textcolor[rgb]{ .2,  .2,  .2}{114.026} & \textcolor[rgb]{ .2,  .2,  .2}{0.4089 } & \textcolor[rgb]{ .2,  .2,  .2}{0.4703 } & \textcolor[rgb]{ .2,  .2,  .2}{208.590 } & \textcolor[rgb]{ .2,  .2,  .2}{0.1669 } & \textcolor[rgb]{ .2,  .2,  .2}{0.2979 } & \textcolor[rgb]{ .2,  .2,  .2}{91.932} \\
    \textcolor[rgb]{ .2,  .2,  .2}{SASRec-full} & \textcolor[rgb]{ .2,  .2,  .2}{× } & \textcolor[rgb]{ .2,  .2,  .2}{0.3948 } & \textcolor[rgb]{ .2,  .2,  .2}{0.4648 } & \textcolor[rgb]{ .2,  .2,  .2}{115.214} & \textcolor[rgb]{ .2,  .2,  .2}{0.4114 } & \textcolor[rgb]{ .2,  .2,  .2}{0.4804 } & \textcolor[rgb]{ .2,  .2,  .2}{211.560 } & \textcolor[rgb]{ .2,  .2,  .2}{0.1666 } & \textcolor[rgb]{ .2,  .2,  .2}{0.3000 } & \textcolor[rgb]{ .2,  .2,  .2}{93.118} \\
    \textcolor[rgb]{ .2,  .2,  .2}{SASRec-small} & \textcolor[rgb]{ .2,  .2,  .2}{× } & \textcolor[rgb]{ .2,  .2,  .2}{0.3856 } & \textcolor[rgb]{ .2,  .2,  .2}{0.4571 } & \textcolor[rgb]{ .2,  .2,  .2}{56.582} & \textcolor[rgb]{ .2,  .2,  .2}{0.4010 } & \textcolor[rgb]{ .2,  .2,  .2}{0.4681 } & \textcolor[rgb]{ .2,  .2,  .2}{68.148 } & \textcolor[rgb]{ .2,  .2,  .2}{0.1628 } & \textcolor[rgb]{ .2,  .2,  .2}{0.2944 } & \textcolor[rgb]{ .2,  .2,  .2}{45.114} \\
    \midrule
    \textcolor[rgb]{ .2,  .2,  .2}{SkipRec} & \textcolor[rgb]{ .2,  .2,  .2}{$\checkmark$ } & \textcolor[rgb]{ .2,  .2,  .2}{0.3940 } & \textcolor[rgb]{ .2,  .2,  .2}{0.4657 } & \textcolor[rgb]{ .2,  .2,  .2}{104.534} & \textcolor[rgb]{ .2,  .2,  .2}{0.4118 } & \textcolor[rgb]{ .2,  .2,  .2}{0.4806 } & \textcolor[rgb]{ .2,  .2,  .2}{197.156} & \textcolor[rgb]{ .2,  .2,  .2}{0.1643 } & \textcolor[rgb]{ .2,  .2,  .2}{0.2966 } & \textcolor[rgb]{ .2,  .2,  .2}{84.926} \\
    \textcolor[rgb]{ .2,  .2,  .2}{AutoDim} & \textcolor[rgb]{ .2,  .2,  .2}{$\checkmark$ } & \textcolor[rgb]{ .2,  .2,  .2}{0.3926 } & \textcolor[rgb]{ .2,  .2,  .2}{0.4610 } & \textcolor[rgb]{ .2,  .2,  .2}{/} & \textcolor[rgb]{ .2,  .2,  .2}{0.4084 } & \textcolor[rgb]{ .2,  .2,  .2}{0.4750 } & \textcolor[rgb]{ .2,  .2,  .2}{/} & \textcolor[rgb]{ .2,  .2,  .2}{0.1649 } & \textcolor[rgb]{ .2,  .2,  .2}{0.2974 } & \textcolor[rgb]{ .2,  .2,  .2}{/} \\
    \textcolor[rgb]{ .2,  .2,  .2}{CANet} & \textcolor[rgb]{ .2,  .2,  .2}{$\checkmark$ } & \textcolor[rgb]{ .2,  .2,  .2}{0.3898 } & \textcolor[rgb]{ .2,  .2,  .2}{0.4587 } & \textcolor[rgb]{ .2,  .2,  .2}{53.226} & \textcolor[rgb]{ .2,  .2,  .2}{0.4045 } & \textcolor[rgb]{ .2,  .2,  .2}{0.4706 } & \textcolor[rgb]{ .2,  .2,  .2}{73.930} & \textcolor[rgb]{ .2,  .2,  .2}{0.1637 } & \textcolor[rgb]{ .2,  .2,  .2}{0.2948 } & \textcolor[rgb]{ .2,  .2,  .2}{37.348} \\
    \midrule
    \multicolumn{2}{c|}{Improve (\%)} & -1.26  & -1.31  & 53.80  & -1.68  & -2.04  & 65.05  & -1.74  & -1.73  & 59.89  \\
    \bottomrule
    \end{tabular}%
  \vspace{-0.15in}
  \label{tab:main}%
\end{table*}}%

\subsubsection{Hyper-parameters and Evaluation Protocols}
For fair comparison purposes, we implement SkipRec and AutoDim on a SASRec backbone same size to CANet. Embedding size is set to 128 for all models. The batchsize for Last.FM20, Last.FM50 and RESS is set to 64, 32, 128 respectively. For optimization, we use Adam optimizer with a learning rate of 0.001 across all experiments.

To quantify the performance of SR models, we evaluate the recommendation results of all models by employing two popular top-$N$ metrics, namely NDCG@$N$ (Normalized Discounted Cumulative Gain), and Recall@$N$. $N$ is set to 10 and 20 for comparison. We apply \textit{leave-one-out} \cite{sasrec,leave_one_out} strategy for evaluation, in which case the evaluated Recall@$N$ is equivalent to HR@$N$. Note that we evaluate each method on the whole item set instead of sampling negative items, which is questioned by \cite{sampled_metric}. To further study the efficiency of each model, we employ FLOPs as another metric. Since CANet and SkipRec are dynamic frameworks, their FLOPs are the average of all inputs. We do not compute the FLOPs of AutoDim since the framework only focuses on the storage efficiency.

\subsection{Performance Analysis}
To save page space, we only report the top-10 results in Table \ref{tab:main}, while the top-20 results follow the similar trend. We also report the relative improvements of CANet compared with SASRec-full in the bottom row of the table. By carefully analyzing the experimental results, we find that deep SR models can better predict user's interests than traditional methods. Besides, self-attentive and convolutional based SR models can be more expressive via stacking deeper hidden layers than shallow GRU4Rec architecture in capturing sequential patterns in user behaviors. Meanwhile, the results also show that larger models usually perform better on SR tasks than smaller ones, which is in consistent with our empirical studies.

Besides, we find that CANet achieves an excellent efficiency on SR tasks. Over all the tested datasets, CANet can reduce an average of 59.2\% computation cost and save up to 65.1\% FLOPs in Last.FM50. Such improvements can greatly reduce the latency in recommendation, ideally with a speedup of more than 2x. On the other hand, the overall degradation on recommendation performance is around  1.5\%, which is basically within the acceptable range. Notably, when compared to SASRec with similar FLOPs, CANet can outperform the baseline model in a wide range. Such results demonstrate that CANet can well reduce the redundancy in large SR models, in the meantime achieves a superior trade-off between efficiency and effectiveness than handcrafted models.

Moreover, CANet can reduce much more FLOPs than SkipRec since the latter only considers the layer depth of network architecture. Such phenomenon again emphasizes the significance of a compound routing space.

\subsection{Ablation Study on Routing Space}
\begin{figure}[htbp]
	\centering
	\includegraphics[width=\linewidth]{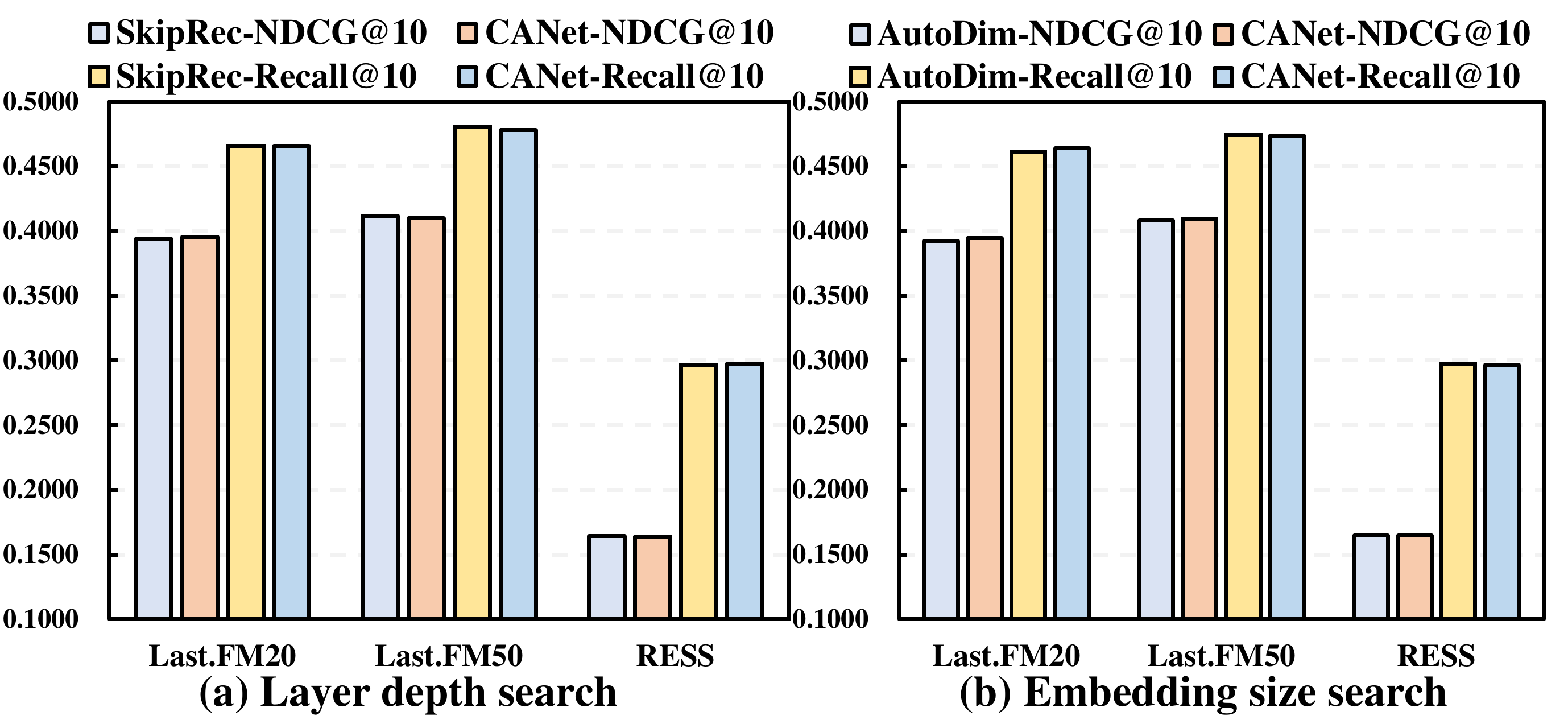}
	\vspace{-0.3in}
	\caption{Ablation study on routing space.}
	\vspace{-0.1in}
	\label{fig:ablation}
\end{figure}

Originally, CANet routes in a compound routing space made up of embedding size, hidden size and layer depth. While we argue that the whole framework has strong scalability, which is competent to tackle difficult heterogeneous searching of compound parameter space but also flexible enough to suit homogeneous tasks like model depth searching. To illustrate our assume as well as for a better comparison with SkipRec and AutoDim, we conduct ablation studies on the routing space of CANet. In detail, we conduct depth searching at an embedding size of 128 compared to SkipRec and embedding size searching at a depth of 8 compared to AutoDim. The searching depth is in the range 1 to 8 while the searching space of embedding size is set to \{64, 96, 128\}. Other hyper-parameters are shared for all models. The results are shown in Figure \ref{fig:ablation}.

Results show that CANet can usually perform competitive results over these two baselines. In some cases, CANet can even perform better in an obvious range, we guess that it is because our proposed Guide loss can help to find submodels with better performance. Note that the baseline models are fixed to handle specific tasks as layer depth searching or embedding size searching respectively, while CANet is capable of a more complex searching task. The comparable results demonstrate CANet's excellent generalization and potential.

\begin{figure*}[htbp]
	\centering
	\includegraphics[width=\linewidth]{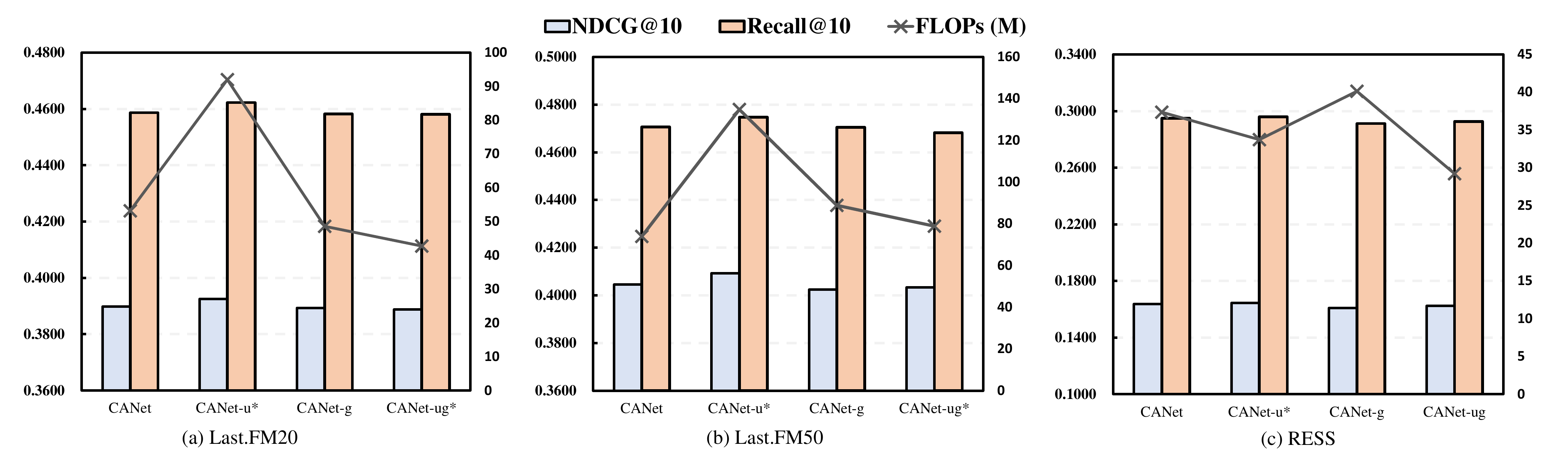}
	\vspace{-0.3in}
	\caption{Test results of CANet and its variants. `*' indicates that the model's router collapsed.}
	\vspace{-0.2in}
	\label{fig:loss}
\end{figure*}

\subsection{Impact of Optimization Strategies}
In this subsection, we turn to analyze the effects of proposed auxiliary loss functions, i.e., the Uniform loss and the Guide loss. We implement three variants of CANet and evaluate their performance.
\begin{itemize}
\item CANet-u, which removes the Uniform loss in optimization objective.
\item CANet-g, which removes the Guide loss in optimization objective.
\item CANet-ug, which removes the Uniform loss and the Guide loss in optimization objective.
\end{itemize}

As illustrated in Figure \ref{fig:loss}, we conclude the findings that: (1) When only utilizing the Gumbel-softmax trick as CANet-ug, model can also achieve high savings in computational cost and maintain overall recommendation results. (2) The model suffers from a collapse issue, which means the output routing signal of the router is a fixed value of one or two. The Uniform loss helps to alleviate the challenge, since both CANet and CANet-g do not encounter such problem. (3) On the other hand, the Guide loss helps the router to discover submodels with better performance. With Guide loss combined, model usually performs better since both CANet and CANet-u outperform their counterparts. These findings support our assumptions that the Uniform loss and the Guide loss can help in the optimization process of CANet.

\begin{figure}[]
	\centering
	\includegraphics[width=\linewidth]{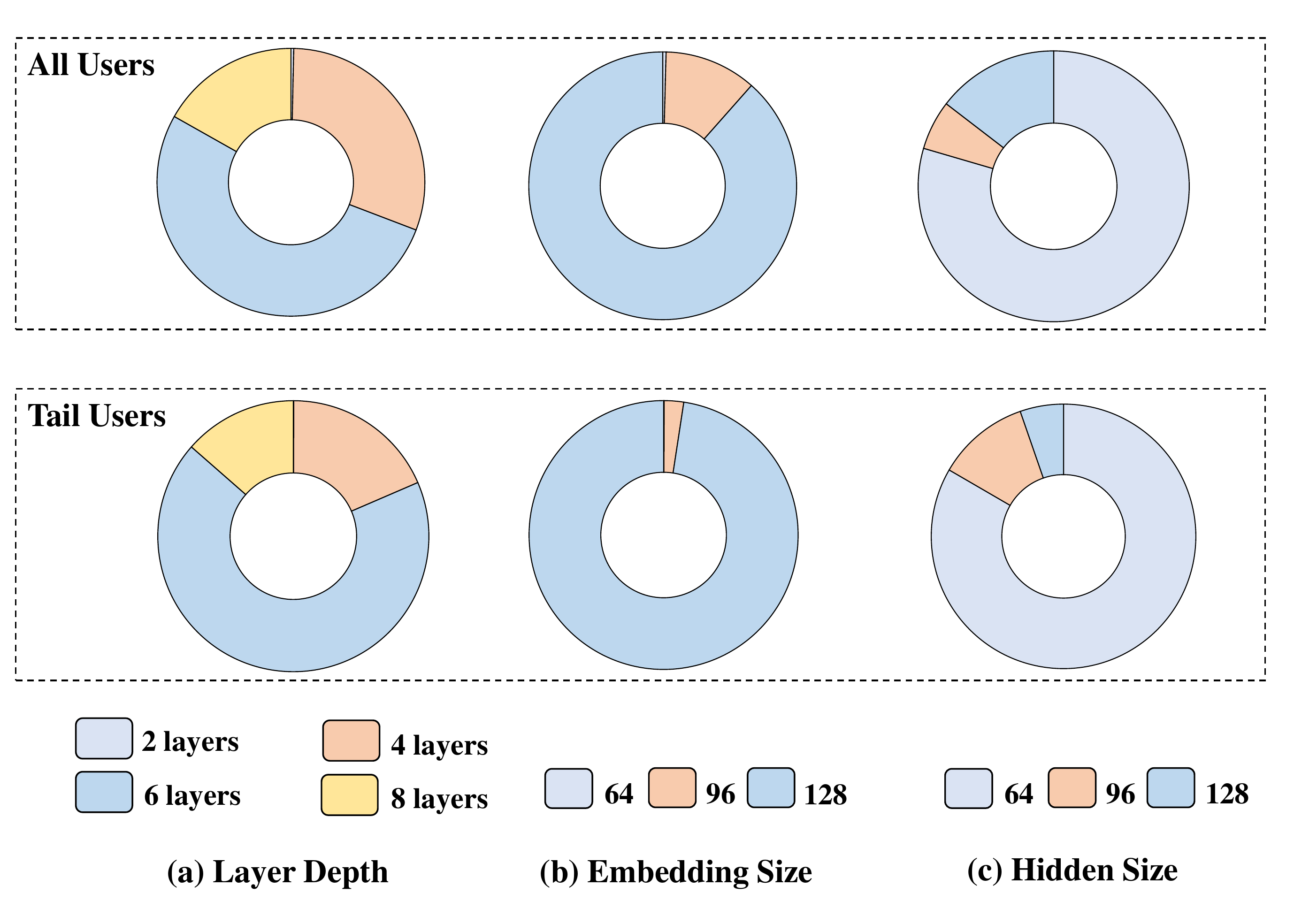}
    \vspace{-0.3in}
	\caption{Routing results of different users on RESS dataset. We count the results from layer depth, embedding size and hidden size respectively.}
	\vspace{-0.2in}
	\label{fig:distribute}
\end{figure}

\subsection{Routing Results Analysis}
To better demonstrate how the router works to reduce redundancy, we visualize the statistics of routing signals. We only report the results on RESS dataset to save page space. Besides, to study whether the router can learn the characteristics of users and thus producing reliable routing signals, we also visualize the statistics of routing signals for tail users. The definition and division of tail user is consistent with section \ref{Empirical Studies}. As illustrated in Figure \ref{fig:distribute}, we can well learn the overall trend of routing results.

In a layer depth view, only few input sequences (0.34\%) are guided to a submodel of 2 middle layers, which indicates that the shallow networks may be not capable enough to model user's evolving preferences. For the rest inputs, the majority of them will pass through a deeper network of 4 or 6 layers. By such early stopping methods, CANet can discover redundancies in model depth in a personalized way and therefore save computation cost adaptively. Furthermore, combining the similar results of Figure \ref{fig:motivation} that SASRec performs well enough with 4 or 6 middle layers, we believe CANet can automatically discover submodels that best balance between computation cost and recommendation performance. 

As for the width of the model, routing signals vary in embedding size and hidden size. It is an interesting phenomenon that most inputs will be first mapped into embedding matrix at a dimension of 128, while the feature extraction process afterwards will be done at a dimension of 64. This may be because that a larger embedding size enables the initial item embedding to carry more information so as to benefit the feature extraction process. While considering that in the scenario of sequential recommendation, all the inputs are discrete IDs without concrete semantics, the non-linear transformation functions in the backbone network may be strong enough to extract user's preference when hidden size is only 64. Consequently, these routing results help to preserve the power of the model while reducing redundancy in width.

Moreover, the comparison of routing results between all users and tail users implies that the router makes decisions on paths according to the user's characteristics. Although the general trend is similar, we notice that there are some notable differences. Specifically, the router assigns deeper models with larger embedding size but smaller hidden size for tail users. Such a difference indicates that the router learns the modeling difficulty of different users and tries to use a stronger model to capture the interests of `hard users'. Meanwhile, the router may also learn the long-tailed characteristics of items, thus using a smaller hidden size to reduce the possibility of overfitting \cite{nis}.

\begin{figure}[htbp]
	\centering
	\includegraphics[width=\linewidth]{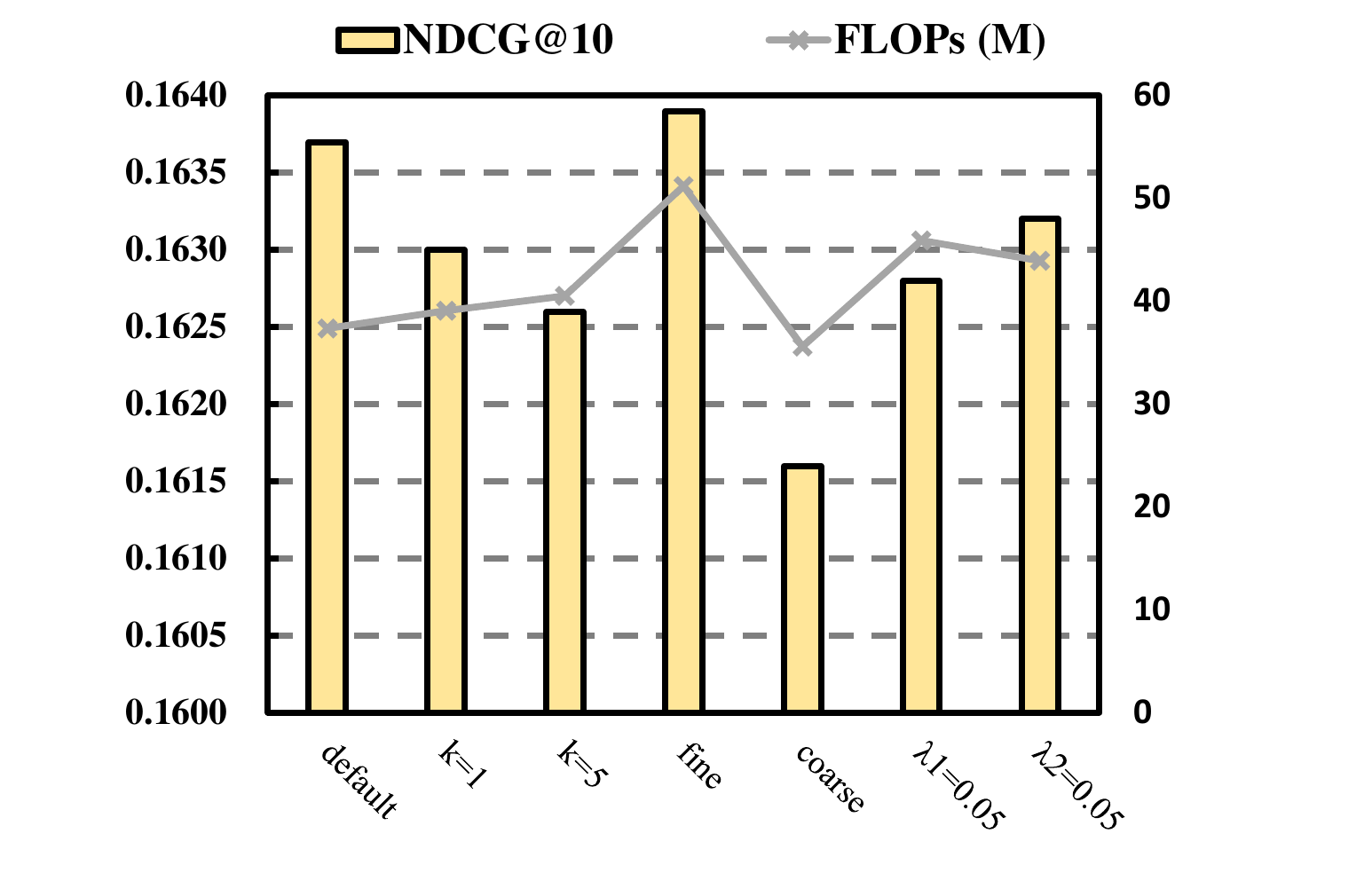}
	\vspace{-0.3in}
	\caption{Hyper-parameter sensitivity study on RESS dataset.}
	\vspace{-0.1in}
	\label{fig:hyper}
\end{figure}

\subsection{Hyper-parameter Sensitivity Study}
In this subsection, we study how the hyper-parameters affect the performance of CANet, including $k$ in the Guide loss, weights $\lambda_1$ and $\lambda_2$ of auxiliary loss functions and granularity of the routing space. Here we further set two routing spaces: We use \textit{fine} to present the routing space of layer in \{1, 2, 3, 4, 5, 6, 7, 8\}, along with embedding size and hidden size in \{64, 80, 96, 112, 128\}. We also construct \textit{coarse} routing space made up of layer in \{2, 8\}, and embedding size and hidden size in \{64, 128\}. We use \textit{default} to present the settings described in the previous sections. Presenting all results on all datasets is redundant and space unacceptable, we mainly report results of hyper-parameter influence on RESS dataset in Figure \ref{fig:hyper}.

We can draw some interesting conclusion based on the results. First, our proposed CANet is slightly sensitive to these hyper-parameters, and the framework can outperform SASRec with similar computation cost under most parameter settings. Such results reflect the robustness of the framework in searching adaptive architecture for modeling user's dynamic preferences. Besides, we also notice that as the routing space granularity becomes finer, CANet can perform better on SR tasks. We leave the application of CANet on more complex routing space for future study.

\section{Related Work}
In this section, we will briefly introduce the related works closely to our work from the following two aspects.

\subsection{Sequential Recommendation}

Sequential recommendation (SR) aims to capture user's dynamic interests to provide recommendations based on their interaction histories. Originally, Markov Chains are first utilized to model user's evolving interests. Utilizing first-order \cite{fpmc} or high-order \cite{highorder} Markov Chain, these models can well capture user's short-term preferences so as to provide more timely and accurate recommendations than traditional methods \cite{bprmf}. With the rise of neural networks, numerous deep learning based methods have emerged. Enhanced by Recurrent neural networks and advanced architectures, models are proven to achieve superior performance \cite{gru4rec, narm, sasrec, nextitnet}. While effective, existing works only focus on the personality of the final representation, in the mean time sharing parameters and model architectures among all users. Consequently, a research point naturally arises that whether a user adaptive model can perform better in sequential recommendation. A recent work \cite{ada} shows the promising results of adaptive parameters and we mainly focus on the personality in network architectures.

\subsection{Efficient Recommendation}

A widely accepted view is that recommendation models are generally huge. Though most previous works only contain very shallow neural networks, the large item set pose significant challenge on both storage and computation efficiency. To tackle such problem, embedding size searching methods \cite{nis,autodim} are widely studied, aiming to search for suitable embedding sizes for input features in an automatic way to reduce parameters in embedding layer. But on the other hand, recent few studies point out that even the hidden layers of recommendation models can also be wider and deeper to gain a performance boost \cite{nasr, skiprec}. How to achieve computation efficient in recommendation is becoming more and more important and some works have been devoted in this searching area. Collaborative distillation \cite{icdm_kd} compresses a pretrained heavy model to a light model to speedup the inference through knowledge distillation, and matrix matrix decomposition methods are also utilized to reduce the computation cost \cite{fan2021lighter, compression}. SkipRec \cite{skiprec} implements a data-dependent pruning framework via layer-skipping methods to select middle layers on a per-user basis. In this way, redundant computation in middle layers can be reduced.

Despite their effectiveness, we hold these methods may lead to a sub-optimal result \cite{dynabert} since they tend to train a tiny but fixed network or implement a single-dimensional adaptive architecture. Different from these works, we aim to further investigate the impact of personality in model architecture in recommendation. To be specific, we propose to equip each person with one adaptive backbone network in a compound view through a dynamic pruning approach \cite{dynamic_survey}, to which we refer as a goal of One person, One model. We verifiy the proposed method in sequential recommendation, and will try its effectiveness in more scenarios later.

\section{Conclusion}

In this work, we presented a focused study on a conceptually simple problem in sequential recommendation, i.e., assigning adaptive network architecture for users in a compound parameter space to reduce computation cost brought by the one-model-fits-all paradigm. We referred to such a goal as One person, One model. We proposed CANet to tackle such challenge, a general and very useful framework to speed the inference of deep SR models. Specifically, we adopted a routing mechanism to adapt each user with personalized network architecture. The proposed CANet showed several properties of being light-weighted, model-agnostic, and flexible for scalability usage. To demonstrate the superiority of our methods, we compared them with competitive baselines. The experimental results showed that our proposed CANet can significantly reduce the computation by $55\sim65\%$ while only sacrificing acceptable performance.

\section{Acknowledgement}
This research was partially supported by grants from the National Key Research and Development Program of China (No.2021YFF0901003), the National Natural Science Foundation of China (Grants No. 61922073, and U20A20229).

\bibliographystyle{IEEEtran}
\bibliography{icdm}
\end{document}